\documentclass{article}

\usepackage{arxiv}

\usepackage[utf8]{inputenc} % allow utf-8 input
\usepackage[T1]{fontenc}    % use 8-bit T1 fonts
\usepackage{hyperref}       % hyperlinks
\usepackage{url}            % simple URL typesetting
\usepackage{booktabs}       % professional-quality tables
\usepackage{amsfonts}       % blackboard math symbols
\usepackage{nicefrac}       % compact symbols for 1/2, etc.
\usepackage{microtype}      % microtypography
\usepackage{lipsum}		% Can be removed after putting your text content
\usepackage{graphicx}
\usepackage{natbib}
\usepackage{doi}
\usepackage{amsmath}
\usepackage{multirow} 
\usepackage{float}
\usepackage{threeparttable}

\title{``The Strength of Weak Ties'' \\ Varies Across Viral Channels \thanks{Thanks to Yifan Yu and Haojun Wu for helpful discussion and support.}}

\author{Shan Huang \thanks{These authors contributed equally to this work.} \hspace{0.1mm} \thanks{To whom correspondence should be addressed. E-mail: shanhh@hku.hk} \\
	The University of Hong Kong \\
	\And
	Yuan Yuan\footnotemark[2] \hspace{0.1mm} \\
	University of California, Davis \\
	\And
	Yi Ji \\
	The University of Hong Kong \\
}

\renewcommand{\shorttitle}{``The Strength of Weak Ties'' Varies Across Viral Channels}

\hypersetup{
pdftitle={A template for the arxiv style},
pdfsubject={q-bio.NC, q-bio.QM},
pdfauthor={David S.~Hippocampus, Elias D.~Striatum},
pdfkeywords={First keyword, Second keyword, More},
}

\begin{document}
\maketitle

\setcounter{footnote}{0} % Reset footnote counter
\renewcommand{\thefootnote}{\arabic{footnote}}

\begin{abstract}
The diffusion of novel information through social networks is essential for dismantling echo chambers and promoting innovation. Our study examines how two major types of viral channels, specifically Direct Messaging (DM) and Broadcasting (BC), impact the well-known ``strength of weak ties'' in disseminating novel information across social networks. We conducted a large-scale empirical analysis, examining the sharing behavior of 500,000 users over a two-month period on a major social media platform. Our results suggest a greater capacity for DM to transmit novel information compared to BC, although DM typically involves stronger ties. Furthermore, the ``strength of weak ties'' is only evident in BC, not in DM where weaker ties do not transmit significantly more novel information. Our mechanism analysis indicates that the content selection by both senders and recipients, contingent on tie strength, contributes to the observed differences between these two channels. These findings expand both our understanding of contemporary weak tie theory and our knowledge of how to disseminate novel information in social networks.
\end{abstract}

% keywords can be removed

\section{Introduction}
Social media networks have emerged as a predominant source for individuals to access novel information~\citep{aral2023exactly, rajkumar2022causal, park2018strength, Bakshy2015}. The exposure to novel information can stimulate creativity and innovation among individuals~\citep{uzzi2013atypical}. It can also help reduce information inequality, bridge societal divides, and encourage active, informed participation in community and societal activities~\citep{Yang2020}. 
Conversely, continuous exposure to homogeneous and redundant information can lead to the fragmentation and polarization of viewpoints. This can exacerbate knowledge divides and deepen societal cleavages~\citep{Pariser2011,tokita2021polarized,sunstein2001republic}. Therefore, understanding how individuals are exposed to novel information through social media networks is a pivotal topic in contemporary social science literature ~\citep{centola2015spreading,rajkumar2022causal,aral2023exactly,jahani2023long}.

The exploration of the role of strong versus weak ties in transmitting novel information has a long-standing history in the literature~\citep{granovetter1973strength,burt2004structural,bakshy2012role, centola2010spread, aral2012identifying, vosoughi2018spread}.
Granovetter's seminal work on the ``strength of weak ties'' emphasizes the unique role of weak ties, which often serve as bridges across different social circles and are essential for disseminating novel and valuable information. In contrast, strong ties often fall short in providing such information. This can be partially attributed to the prevalence of homophily inherent in strong ties, wherein individuals with similar interests and information sources tend to associate together, further fostering similarities~\citep{mcpherson2001birds, centola2015spreading}.
Moreover, having weak ties confers strategic advantages, particularly in accessing a broader spectrum of information and opportunities. For instance, the ``structural hole'' theory introduced by Burt (2004)~\cite{burt2004structural} suggests that individuals bridging different communities through their network positions gain strategic advantages, enabling them to access and disseminate diverse information across those communities.

While social network literature highlights the significance of weak ties in transmitting novel information, debates continue regarding the relative effectiveness of strong versus weak ties as conduits for novel information. There have been studies~\cite{Aral2016, gee2017social, gee2017paradox,rajkumar2022causal} suggesting that strong ties can sometimes play a more crucial role than weak ties in providing novel information.

Additionally, a growing body of research has focused on the factors that moderate the relationship between tie strength and information novelty. For instance, Aral and Van Alstyne (2011)~\cite{aral2011diversity} proposed that the information environment, defined by the speed of updates and diversity of information load, may serve as a significant moderator in the relationship between tie strength and information novelty. This suggests that the effectiveness of strong and weak ties in providing novel information can be context-dependent.

We explore an important yet under-explored factor: the viral channels, or the modes of communication through which information spreads from person to person in social networks. We consider viral channels as a significant moderator on the role of social ties in the transmission of novel information. Specifically, we focus on two distinct major types of viral channels---direct messaging (DM) and broadcasting (BC)---which are prevalently used by social media platforms to disseminate information in their social networks.

Direct messaging, with its inherently private and intimate nature, facilitates a personalized and targeted exchange of information, tailored to the specific needs and interests of the recipients~\cite{treem2012social}. In contrast, broadcasting allows for the dissemination of information to a broad audience, impacting a diverse array of ties—including both strong and weak—without discrimination~\cite{barasch2014broadcasting}. Despite the widespread adoption of DM and BC, the specific roles that these channels play in the dissemination of novel information in social networks are not well understood.

We therefore analyzed a large-scale dataset tracking the information-sharing behaviors of a randomly selected sample of 500,000 users on a major social media platform over two months.\footnote{This platform is one of the largest social media platforms, with over a billion users. Due to a nondisclosure agreement (NDA), we cannot disclose its name.} Our analysis focuses on the dissemination of \textit{online articles} covering a broad range of topics, including but not limited to society, politics, finance, and sports. 
Our analysis examined all the sharing behaviors of online articles by our sampled users during the observation period, recording a total of 567,576 articles shared via DM and 8,718,366 articles disseminated through BC.\footnote{We conducted robustness checks to address the potential concerns of empirical identifications regarding the sample size differences between DM and BC, ensuring the validity of our findings. See \hyperref[SI:robustness]{\textit{SI Appendix C}} for more details.}

Our results demonstrate that, on average, the information transmitted through DM is significantly more novel compared to BC, despite DM involving stronger ties that are conventionally believed to transmit more non-novel information. This trend is consistent across all levels of tie strengths. Furthermore, we observe that in DM, information transmitted by weaker ties is indeed not necessarily more novel than that shared by stronger ties, a phenomenon that diverges from the predictions of conventional weak tie theory or ``the strength of weak ties.'' In contrast, in BC, weaker ties are significantly more effective in disseminating novel information than stronger ties. This divergence underscores the moderating role of viral channels in ``the strength of weak ties'' in transmitting novel information through social networks. 

Our analysis further suggests that content selection by both senders and recipients, contingent on tie strengths, contributes to the differential pattern observed in DM compared to BC. Specifically, our results show that the information shared by senders via DM is significantly less aligned with the senders' historical interests compared to that shared via BC. As a result, in DM, the overlap between the information shared by senders and the interests of their strong-tie recipients can be negatively affected. Additionally, in DM where information can be personalized for different recipients, we observe that senders tend to share more unique and diverse information with their stronger-tie peers than with their weaker-tie peers. This further increases the novelty of information shared through stronger ties compared to weaker ties. On the other hand, in BC, the same information is broadcast indiscriminately to recipients regardless of their social ties with the sender. 
Moreover, we find that recipients in DM tend to engage more with the novel information transmitted through stronger ties, whose endorsement effects are often larger, especially in one-on-one communications like DM. However, such a tendency is not observed in BC, where the broadcasting nature may reduce such endorsement effects. This contributes to the explanation for the higher effectiveness of stronger ties in transmitting novel information in DM compared to BC.

Our findings highlight the interplay of viral channels, tie strength and novel information dissemination in social networks. They suggest that the effectiveness of novel information dissemination is not solely dependent on the strength of ties but is also significantly influenced by the viral channel. These findings add a new dimension to the conventional understanding of the ``strength of weak ties,'' underscoring the need for a more nuanced approach in leveraging social ties and viral channels for effective dissemination of novel information in social networks.

\section{Results}
\label{sec:results}
\subsection{Viral Channels and Information Novelty}
We evaluate the novelty of articles shared via BC and DM on the social media platform under study. Each shared article is referred to as a ``message.'' The novelty level is assessed from the perspective of the information recipients. We measure the average novelty contributed by each social tie, focusing on the following dimensions:
\begin{itemize}
\item \textbf{Marginal Uniqueness:} For a given tie, this refers to the probability that a shared message involves a unique topic or originates from a unique source compared to the topics or sources shared through other ties.
\item \textbf{Marginal Non-redundancy:} For a given tie, this is the probability that a shared message introduces a topic or source distinct from those the recipient has received historically.
\item \textbf{Marginal Diversity:} This indicates how a tie marginally changes the diversity level (topic-wise or source-wise) of messages from all the social ties. Diversity is measured using Shannon entropy, which reflects the variability of information a recipient digests. 
\end{itemize}
Each dimension is analyzed in terms of both topics and sources. ``Topics'' refer to content tags that are generated by the platform's algorithms and assigned to each article. 
Typically, articles are tagged with one topic, though a small fraction may cover two to three tags. On the other hand, ``sources'' refer to the entities that initially posted the articles online, including official media accounts and influencers' accounts. Note that a message is considered \textit{effectively transmitted} in both channels only if the recipient not only receives it, but also \textit{clicks} the link and reads the article. If an article appears in the recipient's BC timeline or DM but is never clicked, it is not deemed as effectively transmitted. The formal definitions of these metrics are discussed in Methods.\footnote{The summary statistics of the metrics are presented in Table~\ref{si:statistics}  in \hyperref[si:statistics]{\textit{SI Appendix A}}.}

%%%%%%%%%%%%%%%%%%%%%%%%%%%%%%%%%%%%%%%%%%%%%%%%%%%%%%%%%%%%%%%%%%
\begin{figure}[H]
\centering
\includegraphics[width=1.0\linewidth]{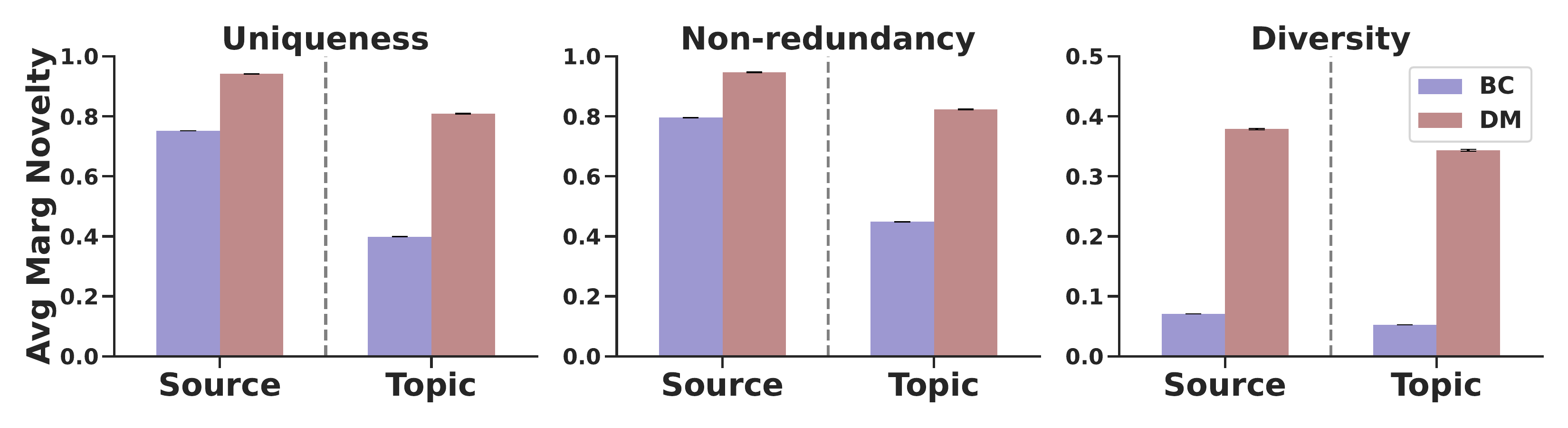}
\caption{Comparison of novelty levels between broadcasting (BC) and direct messaging (DM) using the six novelty metrics. Each bar represents the average novelty per tie for each channel, with error bars delineating the 95\% confidence intervals. Across all metrics, results consistently demonstrate that, on average, the information transmitted through a tie on DM per message is significantly more novel than that transmitted through ties on BC.}
\label{fig:BC_DM_comp}
\end{figure}
%%%%%%%%%%%%%%%%%%%%%%%%%%%%%%%%%%%%%%%%%%%%%%%%%%%%%%%%%%%%%%%%%%

Our findings consistently demonstrate that DM outperforms BC in transmitting novel information across all evaluated novelty metrics, as illustrated in Figure~\ref{fig:BC_DM_comp}. Specifically, the probability of encountering unique information sources and topics is significantly higher through DM ties than BC ties, with a 19.0\% and 41.0\% greater probability than BC ties for topics and sources, respectively ($p<0.01$). 
Regarding non-redundancy, the probability of encountering non-redundant topics and information sources is also greater in DM ties, with a 15.1\% greater probability in non-redundant topics and a 37.5\% greater probability in non-redundant sources ($p<0.01$). Furthermore, when evaluating marginal diversity, DM ties exhibit higher novelty in both source and topic dimensions compared to BC ties, with 0.308 larger Shannon entropy for source-wise novelty and 0.291 larger Shannon entropy for topic-wise novelty ($p<0.01$). 
These metrics collectively support the conclusion that, on average, DM exhibits a higher level of novelty in the information transmitted compared to BC.\footnote{The robustness check results for novelty across channels, presented in Figures~\ref{fig:fig1_robust}~and~\ref{fig:fig1_matching} in \hyperref[SI:robustness]{\textit{SI Appendix C}}, are consistent with the main findings.}

\subsection{Tie Strength and Channel Difference}
To further explore the differences between channels, we next analyze the distributions of tie strength within each channel and examine how tie strength predicts with the transmission of novel information across both channels.
We measure tie strength using the metric of \textit{social embeddedness}, which is quantified by the number of shared social groups (e.g., chat groups with various functions, such as among family members, colleagues, schoolmates, or other social activities in general) between the connected two nodes. Social embeddedness reflects the degree of cohesiveness and overlap between two nodes in a social tie.\footnote{In the context of our study, chat groups that function as online social groups are prevalent on the social media platform we study.} High-embeddedness ties are often characterized by strong bonds, frequent interactions, and shared connections within a closely-knit community, whereas low-embeddedness ties indicate less frequent interactions and connections to different social circles~\citep{Granovetter1985}. 

\begin{figure}[H]
\centering
\includegraphics[width=0.5\linewidth]{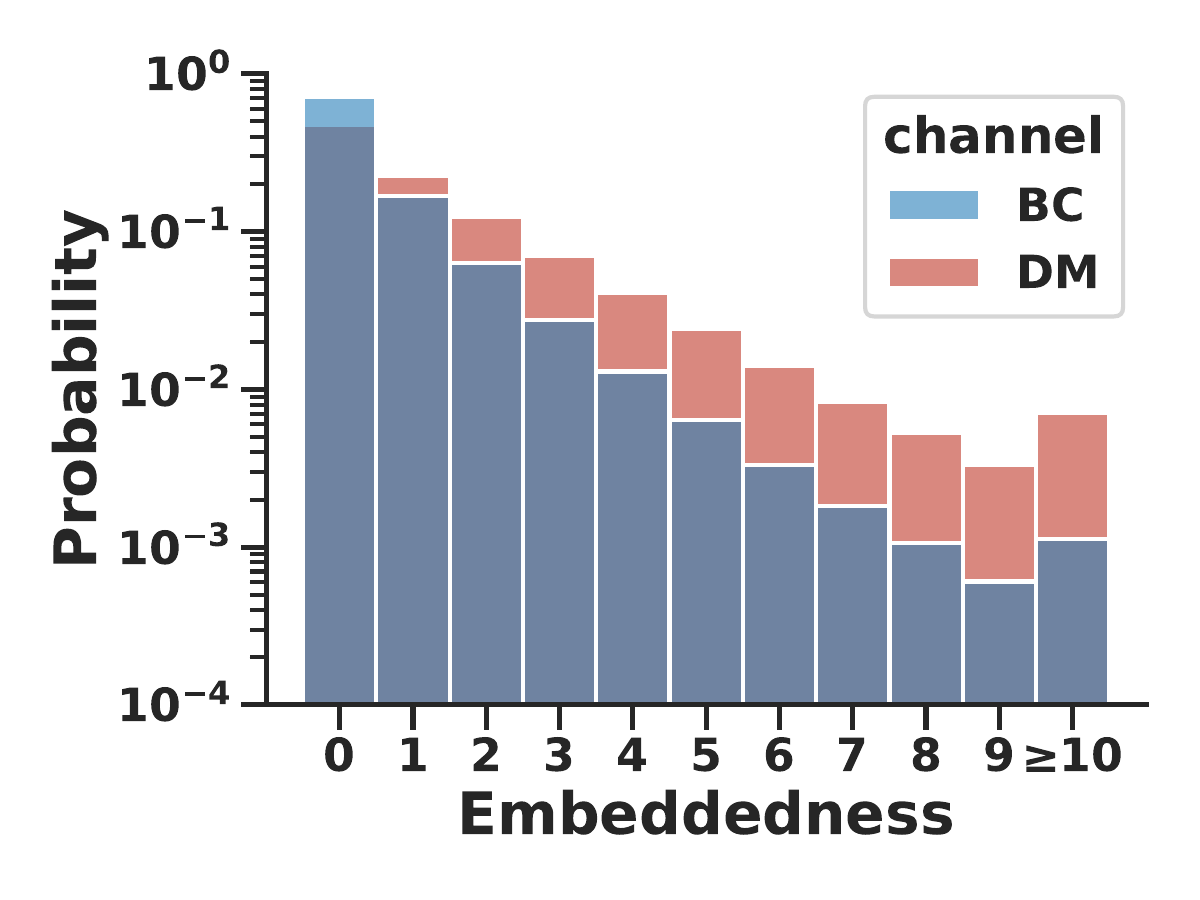}
\caption{Distribution of social embeddedness (measured by the number of mutual chat groups for ties in each channel). For both channels, we examine social ties that shared at least one message through that channel. 
for ties transmitting at least one message, those established through DM generally exhibit greater social embeddedness compared to those through BC.}
\label{fig:BC_DM_band_emb_dist}
\end{figure}

Figure \ref{fig:BC_DM_band_emb_dist} shows that ties in DM demonstrate a larger strength on average, or a higher level of average social embeddedness, compared to those in BC. The average embeddedness of ties in DM is, on average, more than double that of those in BC, a statistically significant difference (\(p < 0.01\)) according to the Kolmogorov-Smirnov (KS) test. 
This result indicates that ties transmitting messages on DM are, on average, more embedded than those on BC. Conventional weak tie theory suggests that stronger ties, which are often characterized by increased similarities among connected nodes, tend to transmit less novel information. However, this does not explain why DM, despite its higher average social embeddedness, is found to transmit more novel information than BC as previously shown in Figure~\ref{fig:BC_DM_comp}. 
This discrepancy suggests that other mechanisms may be influencing the transmission of information in DM in addition to tie strength distributions.\footnote{The robustness check results for distribution of social embeddedness, presented in Figure~\ref{fig:hist_robust} and Figure~\ref{fig:hist_matching} in \hyperref[SI:robustness]{\textit{SI Appendix C}}, are consistent with the main findings.}

%%%%%%%%%%%%%%%%%%%%%%%%%%%%%%%%%%%%%%%%%%%%%%%%%%%%%%%%%%%%%%%%%%%%%%%
%%%%%%%%%%%%%%%%%%%%%%%%%%%%%%%%%%%%%%%%%%%%%%%%%%%%%%%%%%%%%%%%%%%%%%%

\begin{figure}[H]
\centering
\includegraphics[width=\linewidth]{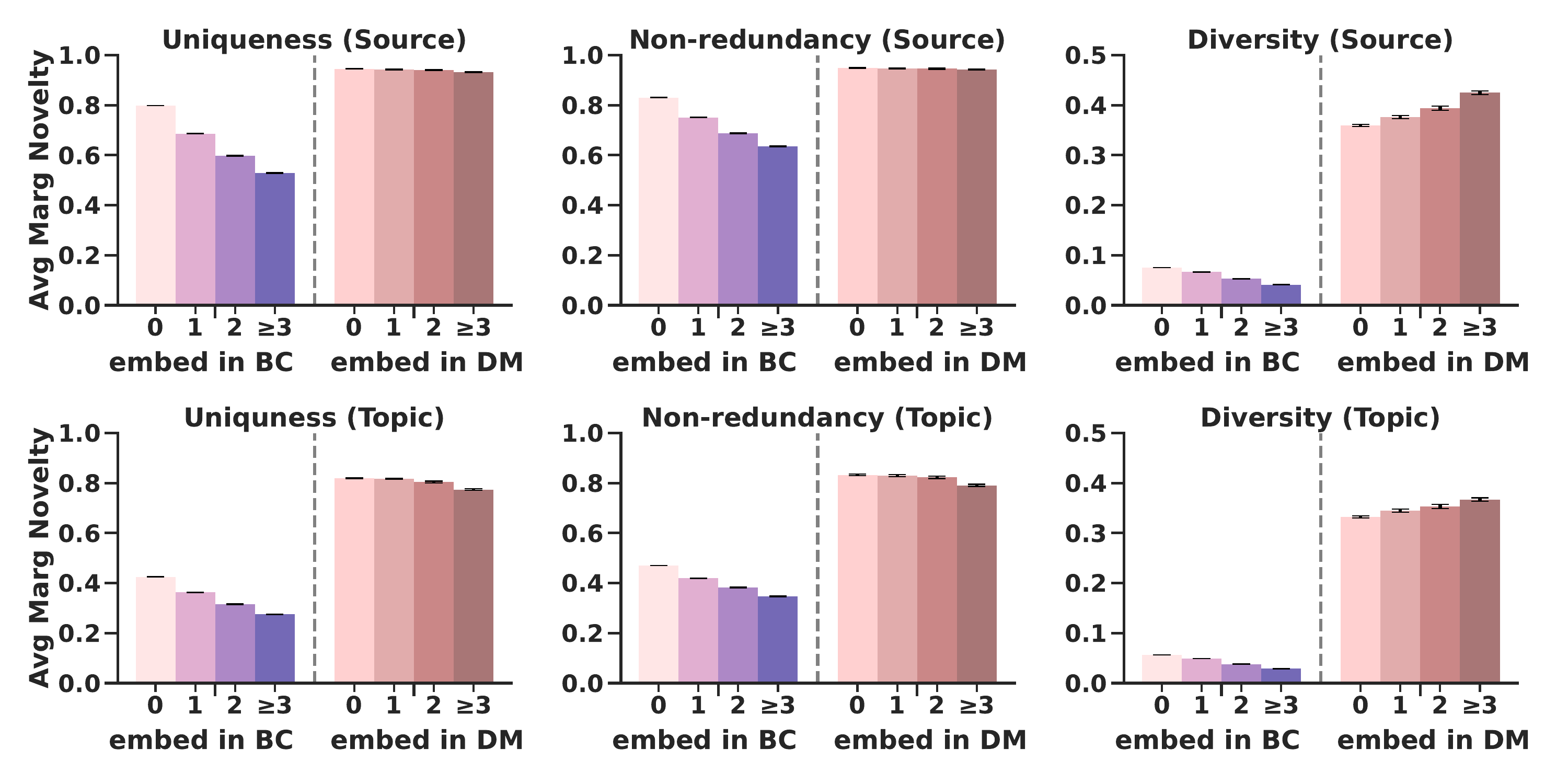}
\caption{The average novelty levels of ties between BC and DM, considering the embeddedness of each tie. The bars represent marginal novelty levels, categorized as either source-wise or topic-wise. Error bars indicate 95\% confidence intervals.}
\label{fig:BC_DM_cond_emb}
\end{figure}

To better understand this discrepancy, we explore the relationship between tie strength and the transmission of novel information through DM versus BC. Specifically, we analyze the average novelty contributed by a social tie conditional on various tie strengths in both channels. As depicted in Figure~\ref{fig:BC_DM_cond_emb}, DM consistently demonstrates a higher level of information novelty compared to BC. This trend holds true for different degrees of embeddedness of ties.

In addition, we observe that in DM, an inverse relationship between tie strength and the novelty of transmitted information is not as evident as in BC. In BC, the results align with the predictions of conventional weak tie theory \citep{granovetter1973strength}, where stronger ties are associated with dissemination of less novel information. Conversely, the observation in DM challenges the result of conventional weak tie theory, as it does not display the expected inverse relationship between tie strength and the novelty of transmitted information. The differences observed between DM and BC channels indicate that the dynamics of information dissemination and novelty vary depending on the communication medium, challenging the generalizability of the ``strength of weak ties'' in the context of diverse viral channels.\footnote{The robustness check results for novelty tendency, presented in Figures~\ref{fig:fig3_robust}~and~\ref{fig:fig3_matching} in \hyperref[SI:robustness]{\textit{SI Appendix C}}, are consistent with the main findings.}

%%%%%%%%%%%%%%%%%%%%%%%%%%%%%%%%%%%%%%%%%%%%%%%%%%%%%%%%%%%%%%%%%%%%%%%
%%%%%%%%%%%%%%%%%%%%%%%%%%%%%%%%%%%%%%%%%%%%%%%%%%%%%%%%%%%%%%%%%%%%%%%

\subsection{Content Selection by Senders and Recipients}
To understand the different patterns observed in DM versus BC, we further investigate potential mechanisms -- content selection by both senders and recipients contingent on social ties. The novelty level of information received by an individual from various social ties is jointly determined by the information that senders choose to share with different tie strengths (i.e., \textit{sender selection effect}) and what recipients decide to engage with from different tie strengths (i.e., \textit{recipient selection effect}). The differences in the selection effects of both senders and recipients between DM and BC may influence the degrees of novelty for individuals received from different social ties across channels. 

\subsubsection{Sender Selection Effect} 
We analyze the novelty level of information that senders choose to share within their local social networks through DM versus BC. As this decision is made by senders, we examine the novelty from senders' perspectives (instead of recipients') to explore the underlying mechanisms. 
 
\begin{figure}[H]
    \centering
    \includegraphics[width=\linewidth]{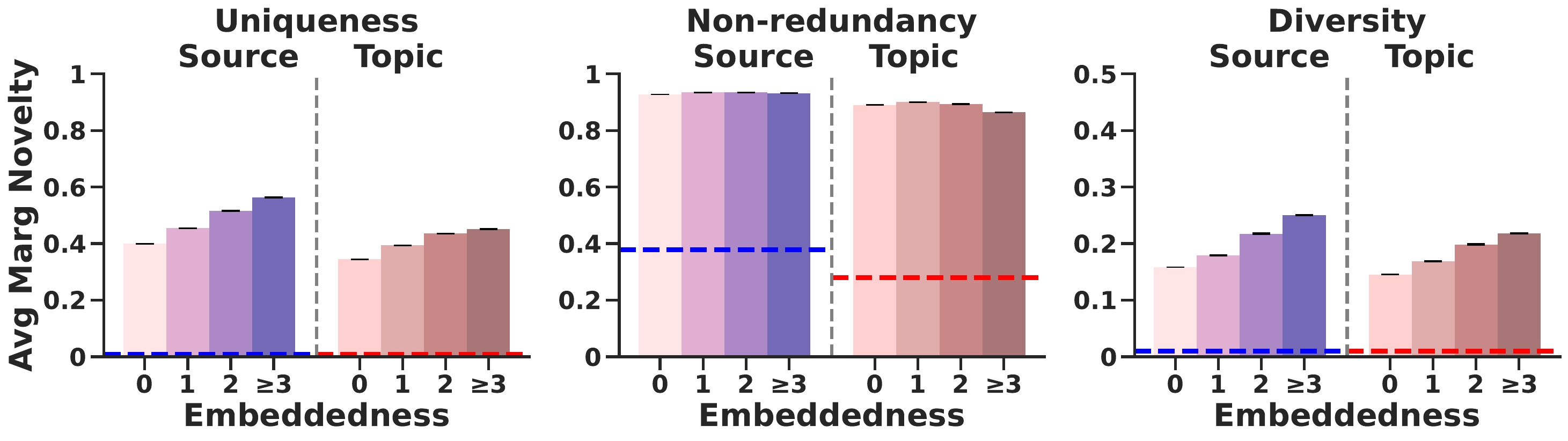}
    \caption{Comparison of average novelty to sender of ties across different tie strengths in DM. The average novelty levels in BC are indicated by the dashed lines. In BC, users share information indiscriminately with friends, resulting in zero values for uniqueness and diversity, and a single value for non-redundancy. Error bars denote 95\% confidence intervals.}
    \label{fig:novelty2sender}
\end{figure}

Figure~\ref{fig:novelty2sender} presents the average novelty to senders in DM and BC across various levels of tie strength. The dashed lines in Figure~\ref{fig:novelty2sender} show the novelty levels to senders in BC, while the bar charts display those in DM. Note that in BC, information is shared indiscriminately and simultaneously with the entire local social network, resulting in no variations in novelty and zero uniqueness and diversity of information across social ties from the senders' perspectives in BC. We thus present results for BC as dashed lines.

For DM, we observed that the non-redundancy level to senders is significantly higher for all levels of tie strength in DM than in BC. This suggests that senders are more inclined to share information that may deviate from their historical interests in DM than in BC. However, there is no apparent trend between tie strength and non-redundancy in the shared information from senders' perspectives. 

One implicit assumption of sharing behavior is that individuals typically share information that reflects their own historical interests. This information tends to align more with the historical interests of strong-tie friends than those of weak-tie friends, due to the potential greater correlated interests among stronger ties. This assumption holds true in BC, as over 60\% of the information shared in BC is redundant and aligns with the senders' historical interests. Consistently, we observe a clear inverse trend between novelty and tie strength in BC in Figure~\ref{fig:BC_DM_cond_emb}.
In contrast, this premise does not hold according to our observations from DM. The information shared in DM shows little overlap with the topics or sources the senders historically consumed. If senders consistently share information not aligned with their own interests, there would not be an evident relationship between tie strength and novelty of information transmitted. 

In addition, Figure~\ref{fig:novelty2sender} shows that senders indeed share unique and diverse information across ties in DM, whereas they disseminate information indiscriminately to their social networks in BC. This result aligns with the distinct mechanism of DM—personalized communication—which is not present in BC. In DM, senders compose customized messages intended for different recipients. As observed previously, these messages may not always align with their own interests, likely driven by the aim to cater to the specific interests of the recipients~\citep{barasch2014broadcasting}. 

Moreover, we note that senders usually share more unique and diverse information with stronger ties, suggesting that closer relationships are associated with more effort to personalize the information shared. This implies that the strength of a tie can positively influence the uniqueness and diversity (novelty) of shared information. 
In conclusion, our findings demonstrate that the senders' ability to personalize content shared with specific recipients in DM, but not in BC, has the potential to reshape the ``strength of weak ties'' in transmitting novelty information. 

\subsubsection{Recipient Selection Effect} 
In addition to senders' selection, the varying tendencies of recipients to engage with information from different social ties can significantly influence how they acquire novel information. Successful transmission of information occurs only when recipients engage with it. We thus investigate the tendency of recipients to selectively engage with novel information, contingent on the strength of the social ties that transmit the information.

We conducted a regression analysis to obtain a deeper understanding of this effect. This analysis examined all messages that were shared but not necessarily clicked by recipients. The dependent variable is whether a recipient clicked on a shared message, considering independent variables including tie strength and the message's novelty (both topic-wise and source-wise) to the recipient. Additionally, we accounted for recipient-level fixed effects. Importantly, our model includes an interaction term between tie strength and novelty, which assists in identifying the recipient selection effect of novelty based on tie strength.

The results are presented in Table \ref{tab:regression:results} in \hyperref[SI:regression]{\textit{SI Appendix B}}. Our findings show that higher-embeddedness ties positively affect message engagement only in DM ($p < 0.01$), but not in BC ($p > 0.1$). Furthermore, recipients are less likely to click  on novel messages in both DM and BC.

Furthermore, the positive and significant interaction effect between tie strength and novelty in DM suggests that recipients are less likely to engage with novel messages if they are sent by less embedded ties ($p<0.01$). This finding implies that, within DM, recipients are more inclined to engage with novel information when shared by stronger-tie senders, as opposed to when the same information is shared by weaker-tie senders. This phenomenon can be attributed to the enhanced endorsement effects of stronger ties in one-on-one DM communications. Strong-tie connections typically exhibit higher levels of trust and shared interests, making recipients more receptive to the novel information they share.

In contrast, such interaction effects are not observed in BC ($p>0.1$). We hypothesize that the difference in interaction effects between DM and BC may be at least partially driven by the fact that, in BC, messages are shared indiscriminately with all of the sender's ties. As such, there is unlikely specific endorsement intention from the sender's perspective towards any particular recipient. Consequently, recipients tend to treat all messages they encounter on their BC timeline more similarly, opting to read messages that seem most interesting and relevant to them, regardless of the specific social ties that shared the message.

Taken together, our observations indicate that in DM, senders selectively share novel information with recipients of different social ties, while such information does not necessarily align with their own interests. From the recipient's perspective, there is a clear bias towards engaging with novel information shared by stronger ties, a phenomenon observed exclusively in DM and not in BC. These results underscore the nuanced nature of information dissemination in DM, offering insights into the counterintuitive relationships between tie strength and information novelty disseminated in this channel and how it differs from BC.

%%%%%%%%%%%%%%%%%%%%%%%%%%%%%%%%%%%%%%%%%%%%%%%%%%%%%%%%%%%%%%%
%%%%%%%%%%%%%%%%%%%%%%%%%%%%%%%%%%%%%%%%%%%%%%%%%%%%%%%%%%%%%%%%

\subsection{Gender Differences} 
\begin{figure}[H]
    \centering
    \includegraphics[width=1.0\linewidth]{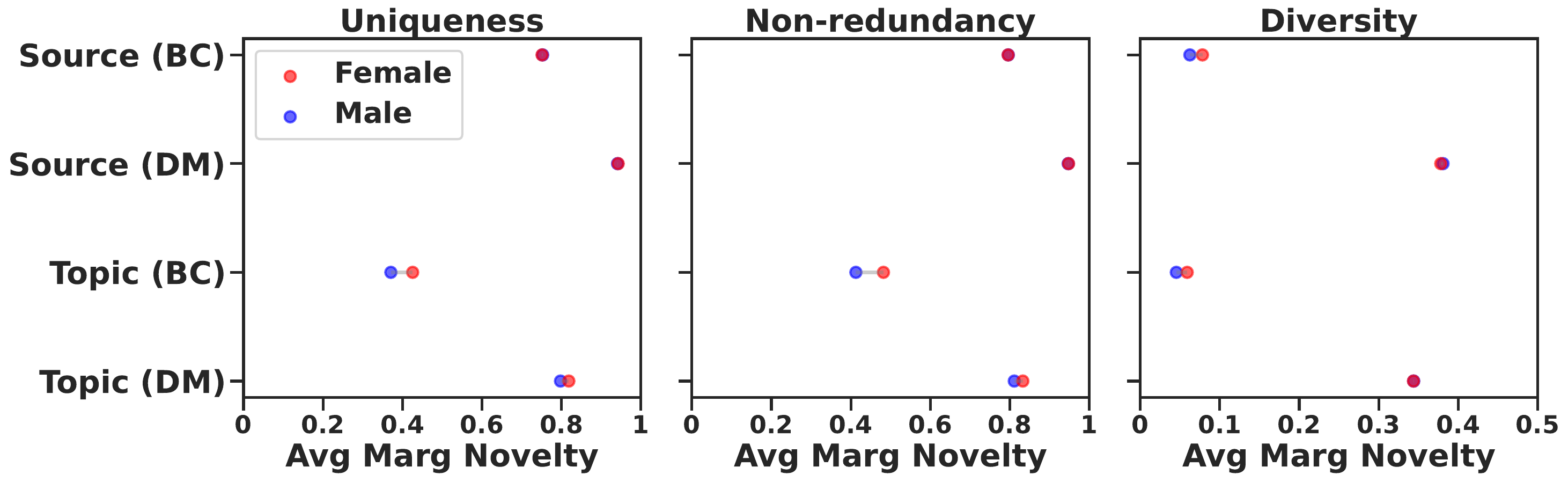}
    \caption{Comparison of average novelty to recipients across genders. The results consistently show that DM delivers significantly more novel information compared to BC for both genders. The findings on gender differences are mixed; some metrics indicate significant differences between females and males, while others do not.}
    \label{fig:gender}
\end{figure}

We further explore gender differences in the exposure of male and female recipients to novel information via DM and BC channels, as illustrated in Figure~\ref{fig:gender}. In this analysis, we separated our sample into female and male groups and reassessed the six novelty metrics across the two channels. Consistently, these results show that DM delivers significantly more novel information compared to BC for both genders. In addition, we observe that the novelty trend across the embeddedness of different genders is consistent with our main analysis, which is illustrated in  Figure~\ref{fig:BC_DM_cond_emb}.

The results concerning gender differences in our study are mixed. For several metrics, such as source-wise uniqueness and non-redundancy, we detect no significant differences between genders. However, for other metrics, specifically topic-wise uniqueness and non-redundancy, we observe that female users receive more novel topics than male users in both channels. This finding aligns with previous literature suggesting that females are more likely to explore new content, a behavior potentially driven by the pleasure derived from reading or a recognition of the connection between reading practices and life success \citep{clark2011young}. Moreover, differences in network structure, communication styles, and social roles may also play a critical role in influencing the diversity of information that genders receive. For example, women may maintain more diverse networks that span both personal and professional contacts, enabling them to access a wider variety of topics from different spheres of life~\cite{diaz2020women}. 

%%%%%%%%%%%%%%%%%%%%%%%%%%%%%%%%%%%%%%%%%%%%%%%%%%%%%%%%%%%%%%%%%%%%%%%%%%%%%%%%%%
%%%%%%%%%%%%%%%%%%%%%%%%%%%%%%%%%%%%%%%%%%%%%%%%%%%%%%%%%%%%%%%%%%%%%%%%%%%%%%%%%

\section{Discussion}
The weak tie hypothesis, which originated from Granovetter's seminal work in 1973~\cite{granovetter1973strength}, remains a cornerstone in discussions about social networks, including topics such as the spread of misinformation, political polarization, and the formation of echo chambers. This hypothesis suggests that weak ties play crucial roles in disseminating novel information, often in critical situations like job searching. Nonetheless, the hypothesis continues to be reexamined, especially as social media becomes an increasingly dominant source of information and large-scale, granular data on individual behaviors become available.

Originally, social media platforms were designed for broadcasting to a wide network. However, more recently, private channels such as direct messaging have also gained popularity, evolving significantly from their non-smartphone origins. Our study investigates the impact of these two communication modes—BC and DM—on the dissemination of novel information. We discovered significant channel differences: DM conveys more novel information than BC, and concerning the disseminating novel information, the distinction between strong and weak ties, as predicted by the conventional weak tie theory~\cite{granovetter1973strength}, is not evident in DM. We explore potential mechanisms from the perspectives of both information senders and recipients to elucidate this channel difference.

Our findings contribute to the ongoing debates regarding the advantages of strong versus weak ties in the diffusion of novel information, emphasizing the crucial role of the viral channel in moderating these relationships. For example, Aral and Van Alstyne (2011)~\cite{aral2011diversity} underscored how factors such as the speed of information change and the complexity of the information environment can serve as significant moderators. 
Nevertheless, our results imply that these factors alone do not entirely account for our observations: since DM and BC employ the same set of information sources (i.e., the same collection of articles disseminated on the same social platform) in our context, the nature of information change or complexity should not differ across channels. This suggests that inherent characteristics of the channels themselves, rather than the nature of the information, are at play. It indicates that the features of the channel, such as potential personalization or endorsement effects in DM and the broad reach nature of BC, may critically impact how novel information is transmitted and shared or received across different viral channels.

Our study presents several intriguing possibilities for future study. First, the direct implications of our findings on critical matters such as misinformation or political polarization necessitate thorough investigation. It is essential to evaluate whether enhancements in a platform's information dissemination mechanisms can promote the spread of novel information into echo chambers, thereby potentially alleviating these issues.
Second, to gain a deeper understanding of the psychological mechanisms underlying channel differences, more research is needed. Specifically, determining the factors that influence whether a person chooses to use DM or BC for information dissemination is vital. This aspect can be examined through meticulously designed surveys or laboratory experiments.

\section{Methods}
\label{sec:methods}
\subsection{Data Collection}
In our study, we randomly sampled 500,000 users from a major social media platform and tracked any articles sent or received through direct messaging (1:1 chats only) or broadcast on users' timelines. For each article, we documented the sender (focal user) and the recipients (their friends), and recorded whether the recipients engaged with the information by clicking the content link. Over the observation period from March 1st to April 30th, 2021, we analyzed a total of 5,341,036 articles. This extensive dataset involved 15,353,612 unique sender-recipient pairs, corresponding to 14,669,698 distinct users and 13,511,473 unique social ties. Importantly, the platform sorts timelines purely by time, thus eliminating potential biases introduced by algorithmic sorting.

To characterize a social tie between two users, we measure two key aspects during our observation period: the number of mutual social groups, which we term ``embeddedness.'' Users build social groups on this social media platform to connect and communicate with others for various reasons, such as groups of colleagues, family, alumni, and groups with mutual interests. Social groups have penetrated every aspect of people's lives and work, and thus can well reflect users' social lives.

We record each article's information sources, the ``official account'' that initially publish the article on the platform, and the topics (content tags) for each article. There are 900,525 unique ``official accounts'' (information sources) in total in our sample. The platform combines manual labeling and machine learning models to ensure that the topics of the articles are accurately annotated, producing 175 topics and each article is associated with 1.07 topics on average. \footnote{The distribution of topics is shown in Table~\ref{si:category-distribution} in \hyperref[si:category-distribution]{\textit{SI Appendix A}}. Additionally, we investigate the transmission pattern of novel information to new audiences, with the results illustrated in Figure~\ref{fig:category} in \hyperref[si:category-hetero]{\textit{SI Appendix D}}.} We also collect data on each user's 5 most frequent topics, generated by the the platform's algorithm with view history data from prior to March 1st. This allows us to measure the novelty of each article in the observation period compared to the user's history.

\subsection{Measuring Embeddedness}
Here, we formally define embeddedness. Let $i$ and $j$ index the individuals, where $\mathcal{N}_i$ indicates the set of users to whom $i$ has sent information. Let $\mathcal{G}_i$ denote the set of chat groups that $i$ belongs to.

We use $\text{em}_{ji}$ to denote the embeddedness, which is measured by the number of mutual groups that both $i$ and $j$ belong to:
\begin{equation}
\text{em}_{ji} = |\mathcal{G}_i \cap \mathcal{G}_j|
\end{equation}

\subsection{Measuring Uniqueness}
Next, we discuss our measure of novelty.
Let $i$ and $j$ be two individuals. 
$\mathcal{N}_i$ denotes the set of neighbors. 
We use $m$ to indicate an article sent from $j$ to $i$. Let $\mathcal{S}$ be the set of information sources (outlets);
we then use $S(m)$ to indicate the source (outlet) of message $m$. Let $\mathcal{T}$ be the set of topics and we use $T(m)$ to indicate the topic of message $m$.\footnote{For simplicity we assume each piece of information has one topic only.} 
By abuse of notation, $S(\mathcal{M}_{ji})$ indicates the set of information sources from $j$ to $i$: 
\begin{equation}
S(\mathcal{M}_{ji}) = \cup_{m \in \mathcal{M}_{ji}} \{S(m)\}
\end{equation}
and $T(\mathcal{M}_{ji})$ indicates the set of topics from $j$ to $i$: 
\begin{equation}
    T(\mathcal{M}_{ji}) = \cup_{m \in \mathcal{M}_{ji}} \{T(m)\}
\end{equation}
The source-wise uniqueness is measured by the number of unique sources shared from $j$ to $i$.
\begin{equation}
\text{num\_uniqueness\_source}_{ji} = \left\lvert S(\mathcal{M}_{ji}) - \cup_{j' \in \mathcal{N}_i \land j' \neq j} S(\mathcal{M}_{j'i}) \right\rvert
\end{equation}
The topic-wise uniqueness is measured by the number of unique topics shared from $j$ to $i$.
\begin{equation}
\text{num\_uniqueness\_topic}_{ji} = \left\lvert T(\mathcal{M}_{ji}) - \cup_{j' \in \mathcal{N}_i \land j' \neq j} T(\mathcal{M}_{j'i}) \right\rvert
\end{equation}
Moreover, we normalize these numbers by the number of messages sent from $j$ to $i$:
\begin{equation}
\text{ratio\_uniqueness\_source}_{ji} = \frac{\text{num\_uniqueness\_source}_{ji}}{|\mathcal{M}_{ji}|}
\end{equation}
Note we use $|\mathcal{M}_{ji}|$ to indicate the number of articles shared from $j$ to $i$. This variable is directional --- $ |\mathcal{M}_{ij}| \neq |\mathcal{M}_{ji}|$.

\begin{equation}
\text{ratio\_uniqueness\_topic}_{ji} =  \frac{\text{num\_uniqueness\_topic}_{ji}}{|\mathcal{M}_{ji}| }
\end{equation}
Note that all the above metrics are asymmetric. Moreover, we employ the ratio rather than the number in our results. 

\subsection{Measuring Diversity}
Diversity is defined on the dyad-level. It means how diverse the messages sent from  $j$  to ego $i$. We use Shannon entropy, a physical property that is most commonly associated with a state of ``randomness'' or ``surprise,'' to define diversity. 

\begin{equation}
    \text{diversity\_source}_{ji} = \sum_{s \in \mathcal{S}} { - q^s_{ji} \log q^s_{ji}}
\end{equation}

\begin{equation}
    \text{diversity\_topic}_{ji} = \sum_{t \in \mathcal{T}} { - q^t_{ji} \log q^t_{ji}}
\end{equation}

\noindent Here  $q^s_{ji}$ and $q^t_{ji}$ are the empirical probability of a message being sent from $j$ to $i$ belonging to the source $s$ and the topic $t$, respectively: $q^s_{ji} = \frac{\left\lvert \left\{ m \in \mathcal{M}_{ji} | S(m)=s \right\}\right\rvert}{\left\lvert \mathcal{M}_{ji} \right\rvert } $ and $q^t_{ji} = \frac{ \left\lvert \left\{  m \in \mathcal{M}_{ji} | T(m)=t \right\}\right\rvert}{\left\lvert\mathcal{M}_{ji}\right\rvert} $. Note that these diversity metrics are heavily affected by the number of messages sent from $j$ to $i$: if there is only one message sent from $j$ to $i$, diversity would be 0; otherwise, the upper bound of diversity would be $\log(|\mathcal{M}_{ji}|)$, which occurs when every message has its own unique source or topic. 

We are interested in the \textit{marginal diversity} $j$ brings to ego, which is defined as 

\begin{equation}
    \text{marginal\_diversity\_source}_{ji} =  \sum_{s \in \mathcal{S}} { - q^s_{i} \log q^s_{i}} - \sum_{s \in \mathcal{S}} { - q^s_{\text{-}j,i} \log q^s_{\text{-}j,i}}
\end{equation}

\begin{equation}
    \text{marginal\_diversity\_topic}_{ji} =  \sum_{s \in \mathcal{T}} { - q^t_{i} \log q^t_{i}} - \sum_{t \in \mathcal{T}} { - q^t_{\text{-}j,i} \log q^t_{\text{-}j,i}}
\end{equation}

\noindent Here $q^s_{i}$ and $q^t_{i}$ are the empirical probability of a message received by $i$ belonging to the source $s$ and the topic $t$, respectively: 
$q^s_{i} = \frac{\left\lvert \left\{ m \in \cup_{j \in \mathcal{N}_i} \mathcal{M}_{{ji}} | S(m)=s \right\}\right\rvert}{\left\lvert \cup_{j \in \mathcal{N}_i} \mathcal{M}_{ji} \right\rvert }$ 
and 
$q^t_{i} = \frac{ \left\lvert \left\{  m \in \cup_{j \in \mathcal{N}_i} \mathcal{M}_{{ji}} | T(m)=t \right\}\right\rvert}{\left\lvert  \cup_{j \in \mathcal{N}_i} \mathcal{M}_{ji} \right\rvert} $. 

\noindent
$q^s_{\text{-}j,i}$ and $q^t_{\text{-}j,i}$ are the empirical probability of a message received by $i$ but not sent by $j$ which belong to the source $s$ and the topic $t$, respectively: 
$q^s_{\text{-}j,i} = \frac{\left\lvert \left\{ m \in \cup_{l \in \mathcal{N}_i \land l \neq j } \mathcal{M}_{li} | S(m)=s \right\}\right\rvert}{\left\lvert \cup_{l \in \mathcal{N}_i \land l \neq j} \mathcal{M}_{li} \right\rvert }$ 
and 
$q^t_{\text{-}j,i} = \frac{ \left\lvert \left\{  m \in \cup_{l \in \mathcal{N}_i \land l \neq j } \mathcal{M}_{li} | T(m)=t \right\}\right\rvert}{\left\lvert  \cup_{l \in \mathcal{N}_i \land l \neq j} \mathcal{M}_{li} \right\rvert} $. 

We use marginal effects to measure the impact of each tie $ji$ in our results.

\subsection{Measuring Non-redundancy}
In addition to a horizontal analysis of novelty measured by uniqueness and diversity during the same period, we also conduct a longitudinal comparison across time. Let $t$ represent the month within the observation period. This period is divided into two sub-periods: $t_{1}$ and $t_{2}$. We use $\mathcal{S}_{i, t_{1}}$ and $\mathcal{T}_{i, t_{1}}$ to indicate the set of information sources and topics received by individual $i$ in the first month. 

For the second month $t_{2}$, we track the sources and topics that individual $i$ encounters. Let $S(\mathcal{M}_{ji, t_2})$ and $T(\mathcal{M}_{ji, t_2})$ denote the set of information—both sources and topics—received by $i$ from $j$ during this month.

Then we compare the information received by users during the second month $t_{2}$ with the historical data in the first month $t_{1}$, and define non-redundancy as follows:

\begin{equation}
\text{NR\_source}_{ji}= | S(\mathcal{M}_{ji, t_{2}}) - \mathcal{S}_{i, t_{1}} |
\end{equation}

\begin{equation}
\text{NR\_topic}_{ji}= |  T(\mathcal{M}_{ji, t_{2}}) - \mathcal{T}_{i, t_{1}}|
\end{equation}

These non-redundancy metrics can also be normalized:

\begin{equation}
\text{ratio\_NR\_source}_{ji}=\frac{\text{NR\_source}_{ji}}{|\mathcal{M}_{ji}|}
\end{equation}

\begin{equation}
\text{ratio\_NR\_topic}_{ji}=\frac{\text{NR\_topic}_{ji}}{|\mathcal{M}_{ji}|}
\end{equation}

We use marginal effects to measure the impact of each tie $ji$ in our results.

\subsection{Analysis}
We first construct an information set for each individual to represent their information consumption profile. This set comprehensively includes all messages received by the focal user. We then evaluate the average tie novelty to the recipient using the six distinct metrics as above: uniqueness, non-redundancy, and diversity, each measured both source-wise and topic-wise. 

For the metrics of uniqueness and diversity, the information set encompasses all messages received by individuals during the entire observation period. Conversely, for the non-redundancy metric, as required by the longitudinal comparative analysis, the information set is limited to articles from the first month of observation. An article is deemed source-wise (topic-wise) unique if the recipient has not concurrently consumed articles from the same source (topic). Source-wise (topic-wise) diversity is calculated by measuring the Shannon entropy difference between the original information set without the articles received through the tie and the augmented set with those articles. An article is considered source-wise (topic-wise) non-redundant to the recipient if its source (topic) does not overlap with the recipient's historically consumed sources (topics) in the last month. 

We normalize the metrics of uniqueness and non-redundancy by dividing them by the number of messages shared via a tie to derive the novelty ratio, which is utilized in further analyses. Next, we categorize ties into various types based on their level of embeddedness (i.e., embed = 0, 1, 2, and $\geq$ 3). For each type, we calculate the average novelty of the tie to the recipient across each communication channel.

The sender-level analysis mirrors the recipient-level analysis, with the exception that all novelty metrics are applied to the senders rather than the recipients. 

\subsection{Sender-level Novelty} 
Here we define the novelty metrics from the perspective of \textit{senders}. Let $i$ be the focal user (sender) and $j$ the recipient. The sender-level source-wise (topic-wise) uniqueness is measured by the ratio of unique sources (topics) shared from $i$ to $j$ compared to $i$'s concurrent sharing in other ties.

\begin{equation}
\text{uniqueness\_source\_sender}_{ij} = \frac{\left\lvert S(\mathcal{M}_{ij}) - \cup_{j' \in \mathcal{N}_i \land j' \neq j} S(\mathcal{M}_{ij'}) \right\rvert}{\left\rvert\mathcal{M}_{ij}\right\rvert}
\end{equation}

\begin{equation}
\text{uniqueness\_topic\_sender}_{ij} = \frac{\left\lvert T(\mathcal{M}_{ij}) - \cup_{j' \in \mathcal{N}_i \land j' \neq j} T(\mathcal{M}_{ij'}) \right\rvert}{\left\rvert\mathcal{M}_{ij}\right\rvert}
\end{equation}

The sender-level source-wise (topic-wise) marginal diversity from $i$ to $j$ is defined as

\begin{equation}
    \text{diversity\_source\_sender}_{ij} =  \sum_{s \in \mathcal{S}} { - q^s_{i} \log q^s_{i}} - \sum_{s \in \mathcal{S}} { - q^s_{i, \text{-}j} \log q^s_{i, \text{-}j}}
\end{equation}

\begin{equation}
    \text{diversity\_topic\_sender}_{ij} =  \sum_{s \in \mathcal{T}} { - q^t_{i} \log q^t_{i}} - \sum_{t \in \mathcal{T}} { - q^t_{i, \text{-}j} \log q^t_{i, \text{-}j}}
\end{equation}

Here, $q^s_{i}$ and $q^t_{i}$ represent the empirical probability of a message sent by $i$ belonging to source $s$ and the topic $t$, respectively: 
$q^s_{i} = \frac{\left\lvert \left\{ m \in \cup_{j \in \mathcal{N}_i} \mathcal{M}_{ij} | S(m)=s \right\}\right\rvert}{\left\lvert \cup_{j \in \mathcal{N}_i} \mathcal{M}_{ij} \right\rvert }$ 
and 
$q^t_{i} = \frac{ \left\lvert \left\{  m \in \cup_{j \in \mathcal{N}_i} \mathcal{M}_{ij} | T(m)=t \right\}\right\rvert}{\left\lvert  \cup_{j \in \mathcal{N}_i} \mathcal{M}_{ij} \right\rvert} $. 

\noindent
$q^s_{i, \text{-}j}$ and $q^t_{i, \text{-}j}$ are the empirical probability of a message sent by $i$ but not received by $j$ which belong to the source $s$ and the topic $t$, respectively: 
$q^s_{i, \text{-}j} = \frac{\left\lvert \left\{ m \in \cup_{l \in \mathcal{N}_i \land l \neq j } \mathcal{M}_{il} | S(m)=s \right\}\right\rvert}{\left\lvert \cup_{l \in \mathcal{N}_i \land l \neq j} \mathcal{M}_{il} \right\rvert }$ 
and 
$q^t_{i, \text{-}j} = \frac{ \left\lvert \left\{  m \in \cup_{l \in \mathcal{N}_i \land l \neq j } \mathcal{M}_{il} | T(m)=t \right\}\right\rvert}{\left\lvert  \cup_{l \in \mathcal{N}_i \land l \neq j} \mathcal{M}_{il} \right\rvert} $. 

For sender-level non-redundancy, we form the same information set defined at the recipient level, representing the information received by the focal user. The sender-level normalized source-wise (topic-wise) non-redundancy is defined as:

\begin{equation}
\text{NR\_source\_sender}_{ij} = \frac{| S(\mathcal{M}_{ij, t_{2}}) - \mathcal{S}_{i, t_{1}} |}{| \mathcal{M}_{ij} |}
\end{equation}

\begin{equation}
\text{NR\_topic\_sender}_{ij} = \frac{|  T(\mathcal{M}_{ij, t_{2}}) - \mathcal{T}_{i, t_{1}} |}{| \mathcal{M}_{ij} |}
\end{equation}

It is important to emphasize that in DM, we follow the definitions and procedures specified above to measure novelty. However, in BC, users disseminate messages to all friends simultaneously, resulting in no novelty differences between different ties. Specifically, for the metrics of uniqueness and diversity, since messages are delivered to friends without differentiation, the values for these two measures of novelty should naturally be zero. Regarding non-redundancy, a single value is calculated to measure the non-redundancy of information sent by users. This is because the information transmitted through ties between a sender and all their friends is identical.

\bibliographystyle{unsrtnat}
\bibliography{references}

\newpage

\section*{Supplementary Information}

\setcounter{section}{0}
\renewcommand{\thesection}{\Alph{section}}
\renewcommand{\thesubsection}{\thesection.\arabic{subsection}}

\setcounter{table}{0}
\setcounter{figure}{0}

\renewcommand{\thetable}{S\arabic{table}}
\renewcommand{\thefigure}{S\arabic{figure}}

\section{Summary Statistics}
We present the summary statistics in Table~\ref{si:statistics}. The distribution of embeddedness aligns with the histogram presented in the main text. For all six novelty metrics, DM exhibits a higher average level than BC, echoing the findings in the main text. It is worth noting that the maximum value for the ratio of uniqueness and non-redundancy should be 1. Due to its definition as the difference in Shannon entropy, the maximum value for marginal diversity can exceed 1. While the minimum value for marginal diversity can be negative, it is typically non-negative in our observations.

We also present the distribution of categories in Table~\ref{si:category-distribution}, ordered by frequency in descending order. Each article is classified by the platform into a broad category and a detailed topic. Our dataset comprises 21 categories and 175 topics. For instance, the category ``culture'' includes topics such as literature, history, and drama. As indicated in the table, the top three categories across both channels are ``society,'' ``education,'' and ``health,'' signifying a strong focus on these areas. Conversely, categories like ``sports,'' ``technology,'' ``fashion,'' and ``entertainment'' occupy smaller fractions, likely due to their niche market appeal.

\begin{table*}[h!]
\centering
\caption{Summary Statistics for Ties}
\label{si:statistics}
\hspace*{-1.2cm}
% \begin{tabular}{cccccccccccccccc}
\begin{tabular}{p{3.75cm}p{0.5cm}p{0.7cm}p{0.7cm}p{0.9cm}p{0.5cm}p{0.5cm}p{0.5cm}cp{0.5cm}p{0.7cm}p{0.6cm}p{0.9cm}p{0.5cm}p{0.5cm}p{0.5cm}}
\toprule
 & \multicolumn{7}{c}{\textbf{Broadcasting}} && \multicolumn{7}{c}{\textbf{Direct Messaging}} \\
\cmidrule{2-8} \cmidrule{10-16}
\textbf{Variable} & \textbf{Max} & \textbf{Min} & \textbf{25\%} & \textbf{Median} & \textbf{75\%} & \textbf{Mean} & \textbf{SD} && \textbf{Max} & \textbf{Min} & \textbf{25\%} & \textbf{Median} & \textbf{75\%} & \textbf{Mean} & \textbf{SD} \\
\midrule
\# of Messages & 930 & 1 & 1 & 1 & 1 & 1.68 & 3.39 && 731 & 1 & 1 & 1 & 2 & 1.74 & 3.39 \\
Embeddedness & 99 & 0 & 0 & 0 & 1 & 0.52 & 1.14 & & 99 & 0 & 0 & 1 & 2 & 1.27 & 1.98 \\
Marginal Uniqueness {\tiny(Source)} & 1 & 0 & 0.5 & 1 & 1 & 0.75 & 0.40 & & 1 & 0 & 1 & 1 & 1 & 0.94 & 0.19 \\
Marginal Uniqueness {\tiny(Topic)} & 1 & 0 & 0 & 1 & 1 & 0.38 & 0.49 & & 1 & 0 & 0.75 & 1 & 1 & 0.85 & 0.45 \\
Marginal Diversity {\tiny(Source)} & 4.97 & -2.60 & 0.01 & 0.04 & 0.09 & 0.07 & 0.14 & & 4.84 & -1.43 & 0 & 0.29 & 0.69 & 0.28 & 0.42 \\
Marginal Diversity {\tiny(Topic)} & 3.51 & -1.69 & -0.01 & 0.01 & 0.07 & 0.05 & 0.14 & & 3.68 & -1.43 & 0 & 0.22 & 0.69 & 0.34 & 0.42 \\
Marginal NR {\tiny(Source)} & 1 & 0 & 1 & 1 & 1 & 0.80 & 0.39 &  & 1 & 0 & 1 & 1& 1 & 0.95 & 0.21 \\
Marginal NR {\tiny(Topic)} & 1 & 0 & 0 & 0 & 1 & 0.45 & 0.48 &  & 1 & 0 & 1 & 1 & 1 & 0.82 & 0.36 \\
\bottomrule
\end{tabular}
\end{table*}

\begin{table*}[h!]
\centering
\caption{Distribution of Categories}
\label{si:category-distribution}
\begin{tabular}{ccccccc}
\toprule
\multicolumn{3}{c}{\textbf{Broadcasting}} & & \multicolumn{3}{c}{\textbf{Direct Messaging}} \\
\cmidrule{1-3} \cmidrule{5-7} 
\textbf{Category} & \textbf{Number} & \textbf{Proportion} & &  \textbf{Category} & \textbf{Number} & \textbf{Proportion} \\\cmidrule{1-3} \cmidrule{5-7} 
Society & 2464777 & 26.42\%  &  & Society & 120652 & 19.65\% \\
Education & 1045294 & 11.20\% & &  Health & 66244 & 10.79\% \\
Culture & 925588 & 9.92\% & & Education &  63641 & 10.37\% \\
Health & 873341 & 9.36\% & & Culture &   54858 & 8.94\% \\
Investment & 475022 & 5.09\% & &  Workplace & 47651 & 7.76\% \\
Workplace & 458043 & 4.91\% &  & Investment & 36534 & 5.95\% \\
Hobby & 453403 & 4.86\% &  & Hobby & 31150 & 5.07\% \\
Travel & 404614 & 4.34\% & &  Travel & 29896 & 4.87\% \\
Lifestyle & 392733 & 4.21\% &  & Lifestyle & 25260 & 4.11\% \\
International & 363580 & 3.90\% & &  International & 22925 & 3.73\% \\
Internet & 226554 & 2.43\% &  & Catering & 20817 & 3.39\% \\
Catering & 202868 & 2.17\% &  & Internet & 13165 & 2.14\% \\
Advertisement & 194218 & 2.08\% & &  Religion & 13140 & 2.14\% \\
Religion & 147059 & 1.58\% &  & Relationship & 12581 & 2.05\% \\
Science & 133433 & 1.43\% &  & Advertisement & 12412 & 2.02\% \\
Relationship & 117692 & 1.26\% & &  Showbiz & 10988 & 1.79\% \\
Showbiz & 114324 & 1.23\% &  & Science & 7394 & 1.20\% \\
Technology & 103686 & 1.11\% &  & Entertainment & 6773 & 1.10\% \\
Sports & 101603 & 1.09\% &  & Fashion & 6430 & 1.05\% \\
Fashion & 80434 & 0.86\% &  & Technology & 5959 & 0.97\% \\
Entertainment & 52239 & 0.56\% &  & Sports & 5472 & 0.89\% \\
\bottomrule
\end{tabular}
\end{table*}

\section{Regression Analysis on Recipients' Selection Effects}
\label{SI:regression}
We performed regression analyses at the message level to explore how recipients engage with messages (articles), considering the tie strength with the sender and the information novelty. Specifically, we analyzed the sample of articles that have been shared to each recipient and examine how the recipient engage with these articles. 
In this analysis, we employed message-level non-redundancy — the degree of novelty compared to an individual's historical consumption pattern — as the measure of novelty. Specifically, the topic-wise (source-wise) novelty is considered one if the message does not contain content associated with the same topic (source) that the recipient has engaged with historically.\footnote{Uniqueness and diversity are measured relative to the information shared by other social ties and, as a result, cannot be measured at the message level. In contrast, non-redundancy is measured relative to the information that recipients have historically received, allowing it to be assessed at the message level for a specific recipient.}

$$\text{logit}(\text{click}_{mi}) = \beta_0 + \beta_1 \times t_{mi} + \beta_2 \times w_{ji} + \beta_3 \times t_{mi} \times w_{ji} + \text{FE}_{i} $$

$$\text{logit}(\text{click}_{mi}) = \beta_0 + \beta_1 \times s_{mi} + \beta_2 \times w_{ji} + \beta_3 \times s_{mi} \times w_{ji} + \text{FE}_{i}$$

For each message (article), denoted by $m$, we investigate how the tie strength between the recipient $i$ and the sender $j$, as well as the novelty of $m$, influence the likelihood of the recipient $i$ clicking on the message $m$, and especially the interaction effects between the tie strength and the novelty of article. 
We use $t_{mi}$ and $s_{mi}$ to represent the novelty of an article for the recipient in terms of topic and source, respectively. The binary variable $w_{ji}$ denotes the tie strength (i.e., social embeddedness) between sender $j$ and recipient $i$. The outcome $\text{click}_{mi}$ indicates whether recipient $i$ clicks on article $m$. 
We run the following logistic regressions and report the average marginal effects. We've included the recipient $i$'s fixed effects ($\text{FE}_i$) to the regression, which controls for individual variation in the tendency of clicking on a message they receive. 
The coefficients $\beta_1, \beta_2, \text{ and } \beta_3$ capture the effects of novelty, tie strength, and their interactions on the likelihood of clicking an article. 

\begin{table*}[h!]
\centering
\caption{Regression Results for Novel Topics and Novel Sources}
\label{tab:regression:results}
\hspace*{-0.7cm}
\begin{tabular}{cccccccc}
\toprule
& 
& \multicolumn{3}{c}{\textbf{Broadcasting} (n = 178,549)}
& \multicolumn{3}{c}{\textbf{Direct Messaging}  (n = 790,372)} 
\\
\cmidrule(lr){3-5} \cmidrule(lr){6-8}
\textbf{Outcome} & \textbf{Variable} & \textbf{Coefficient} & \textbf{Std. Error} & \textbf{p-value} & \textbf{Coefficient} & \textbf{Std. Error} & \textbf{p-value} \\
\midrule
\multirow{1}{*}{Click} 
& Novel Topic & -0.0132 & 0.0015 & < 0.01 & -0.0259 & 0.0012 & < 0.01 \\
\midrule
\multirow{1}{*}{Click} 
& Novel Source & -0.0053 & 0.0020 & < 0.01 & -0.0501 & 0.0018 & < 0.01 \\
\midrule
\multirow{1}{*}{Click} 
& Embeddedness & 0.0021 & 0.0024 & 0.382 & 0.0100 & 0.0003 & < 0.01 \\
\midrule
\multirow{3}{*}{Click} 
& Novel Topic & -0.0138 & 0.0017 & < 0.01 & -0.0285 & 0.0014 & < 0.01 \\
& Embeddedness & 0.0018 & 0.0024 & 0.461 & 0.0081 & 0.0005 & < 0.01 \\
& Novel Topic $\times$ Embeddedness & 0.0010 & 0.0013 & 0.424 & 0.0026 & 0.0005 & < 0.01 \\
\midrule
\multirow{3}{*}{Click} 
& Novel Source & -0.0057 & 0.0023 & 0.016 & -0.0576 & 0.0020 & < 0.01 \\
& Embeddedness & 0.0015 & 0.0025 & 0.531 & 0.0046 & 0.0007 & < 0.01 \\
& Novel Source $\times$ Embeddedness & 0.0005 & 0.0015 & 0.716 & 0.0057 & 0.0007 & < 0.01 \\
\bottomrule
\end{tabular}
\begin{tablenotes}
\footnotesize
\item Note: The source-wise (topic-wise) novelty is calculated by comparing users' article reception records in the second month with their historical engagement in the first month. Therefore, the sample size for regression is around half of the total observations for the entire period.
\end{tablenotes}
\end{table*}

\noindent\textit{Sampling:} Identifying all the information that appears in recipients' timelines can be challenging; therefore, we could only collect data on articles that were clicked and effectively received in BC. This leaves us uninformed about messages that were shared but not clicked for recipients in BC. However, for DM, we are able to collect data on both shared and clicked messages.
We, therefore, restructured the BC data by focusing on ties (denoted by $(j,i)$) that exist in both DM and BC. Specifically, if we identified a message $m$ shared by a sender $j$ in BC that was not clicked by their peer $i$, we added the tuple $(m,i,j)$ to the sample with the outcome variable set to 0. This approach enabled us to reconstruct a sample to analyze click behavior comparably for both BC and DM. As a result, we examined 17,363 ties and 370,178 observations (tuples of $(m,i,j)$) to obtain our findings, thereby ensuring a more balanced comparison between BC and DM.\footnote{Note that there is a small sample bias induced --- we now focus on ties that have shared (but not necessarily effectively transmitted or clicked) at least one message through DM for both channels.} 

\noindent\textit{Results:} From Table~\ref{tab:regression:results}, we observe the positive interaction effects between embeddedness and information novelty consistently in DM, indicating that people are more receptive to novel information from stronger ties in DM. However, in BC, the interaction term between novelty and weak ties is much smaller and non-significant ($p > 0.1$).
These results suggest that in BC, the endorsement effect of strong ties on novel information is absent. We propose that, unlike in DM, in BC, senders distribute messages indiscriminately to all recipients without engaging in one-on-one communications, which could potentially dilute the endorsement effect. 

Overall, these results highlight a significant recipient selection effect favoring novel information from stronger ties in DM, an effect that is not present in BC. In DM, recipients tend to be more receptive to novel messages shared by their stronger ties. In contrast, in BC, recipients do not exhibit any significant differences in their receptiveness to novel information shared by strong versus weak ties. The disparity in recipient selection effects between DM and BC also partially explains the channel differences concerning the less significant ``strength of weak ties'' in DM than in BC, as observed in Figure~\ref{fig:BC_DM_cond_emb}.
 
\section{Robustness Checks}
\label{SI:robustness}
As we analyzed field data, DM and BC naturally have sample differences. In this section, we conducted robustness checks to demonstrate that our findings are not influenced by these sample differences. First, we focus on recipients who appear in both DM and BC and replicate our analysis using the same sample for both viral channels. In the second robustness check, we performed matching at the recipient level based on their bandwidth --- the total number of received messages.

\subsection{Common Users in DM and BC}
\begin{figure}[H]
    \centering
    \includegraphics[width=0.7\linewidth]{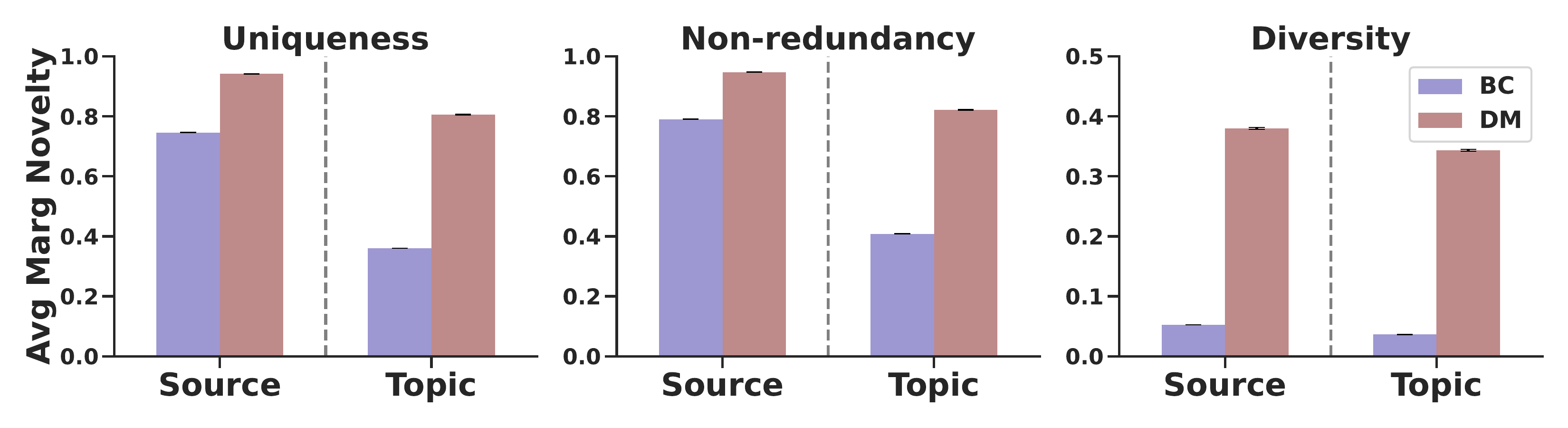}
    \caption{Robustness check for common recipients: Comparison of novelty levels between broadcasting (BC) and direct messaging (DM) using the six novelty metrics. Each bar graph represents the average novelty per tie and message for each channel, with error bars delineating the 95\% confidence intervals. Across all metrics, results consistently demonstrate that, on average, the information transmitted through a tie on DM per message is significantly more novel than that transmitted through ties on BC. The results are consistent with the main text.}
    \label{fig:fig1_robust}
\end{figure}

\begin{figure}[h!]
    \centering
    \includegraphics[width=0.4\linewidth]{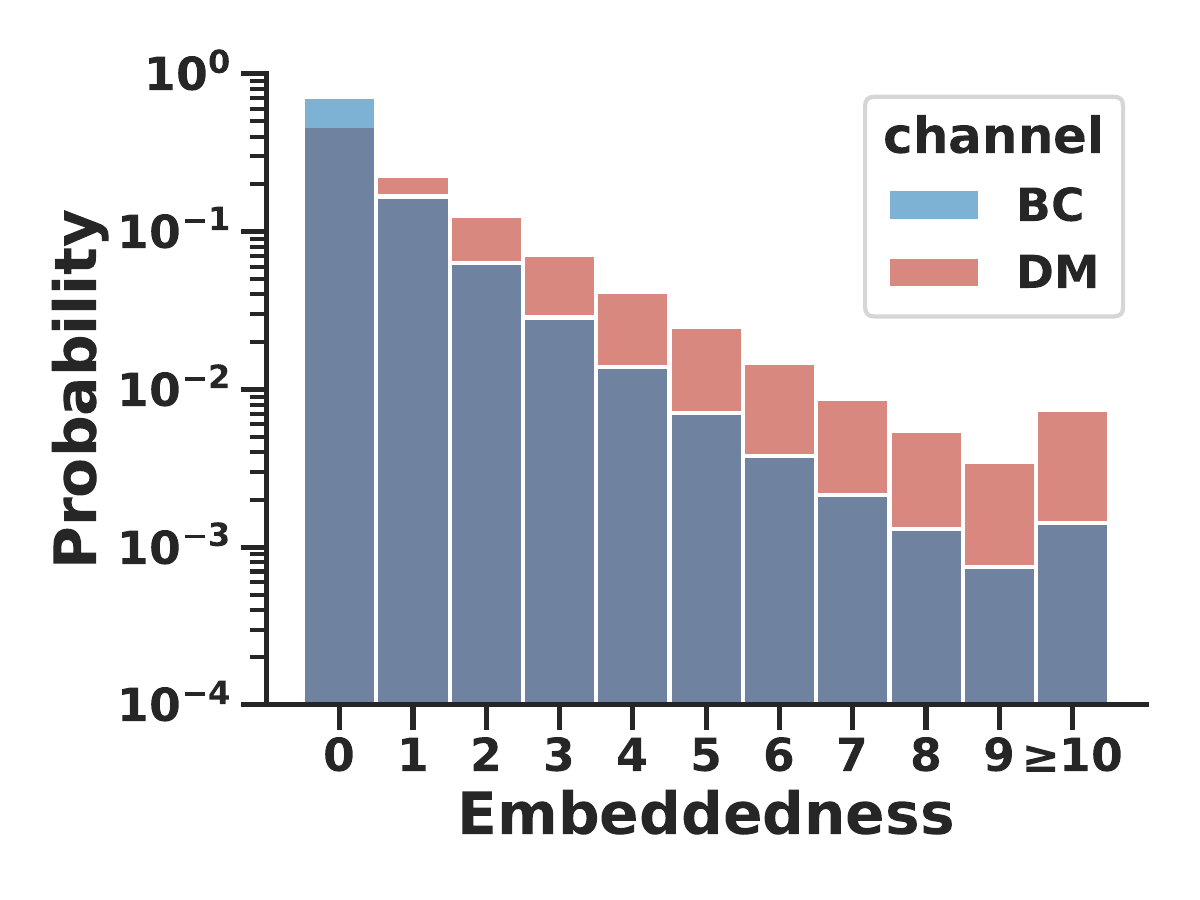}
    \caption{Robustness check for common recipients: Distribution of social embeddedness (measured by the number of mutual chat groups for ties in each channel). For both BC and DM, we examine social ties that shared at least one message through that channel. The results are consistent with the main text.}
    \label{fig:hist_robust}
\end{figure}

\begin{figure}[h!]
    \centering
    \includegraphics[width=0.7\linewidth]{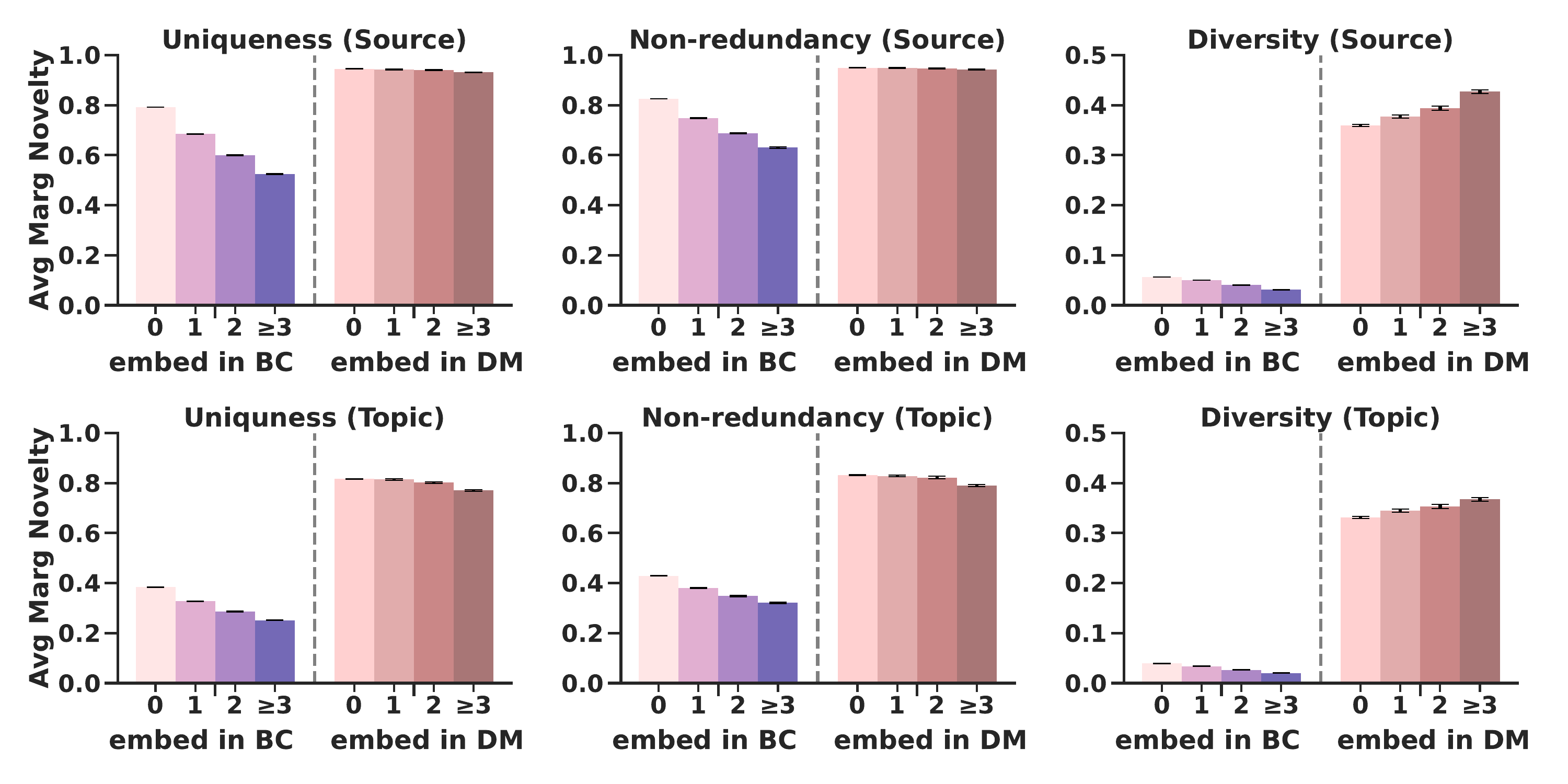}
    \caption{Robustness check for common recipients: Comparison of average novelty levels of ties in BC and DM, conditional on the embeddedness of the tie. Error bars denote 95\% confidence intervals. The results are consistent with the main text.}
    \label{fig:fig3_robust}
\end{figure}

To further verify the robustness of our findings, we focus on recipients who have received (clicked on) at least one message from both channels. We therefore analyze a subsample of 170,712 recipients and replicate the main analysis. The results, presented in Figures \ref{fig:fig1_robust} to \ref{fig:fig3_robust}, further validate the main findings of our study, demonstrating the consistency of the observed patterns.

%%%%%%%%%%%%%%%%%%%%%%%%%%%%%%%%%%%%%%%%%%%%%%%%%%%%%%%%%%%%%%%%%%
%%%%%%%%%%%%%%%%%%%%%%%%%%%%%%%%%%%%%%%%%%%%%%%%%%%%%%%%%%%%%%%%%%

%%%%%%%%%%%%%%%%%%%%%%%%%%%%%%%%%%%%%%%%%%%%%%%%%%%%%%%%%%%%%%%%%%
%%%%%%%%%%%%%%%%%%%%%%%%%%%%%%%%%%%%%%%%%%%%%%%%%%%%%%%%%%%%%%%%%%

%%%%%%%%%%%%%%%%%%%%%%%%%%%%%%%%%%%%%%%%%%%%%%%%%%%%%%%%%%%%%%%%%%
%%%%%%%%%%%%%%%%%%%%%%%%%%%%%%%%%%%%%%%%%%%%%%%%%%%%%%%%%%%%%%%%%%

\subsection{Matching Based on Bandwidth}
We observe that, compared to those in BC, the recipients in DM exhibit a lower bandwidth.\footnote{We define the bandwidth of a recipient as the total number of received messages (i.e., $bw_{i} = \sum_{j\in{N_{i}}}{|\mathcal{M}_{ji}|}$).} The average bandwidth recorded was 3.04 for DM and 20.53 for BC. This apparent discrepancy could contribute to a higher average marginal novelty in DM, given that its recipients have a smaller information set to compare. This is especially relevant in terms of uniqueness and diversity.

\begin{figure}[H]
    \centering
    \includegraphics[width=0.7\linewidth]{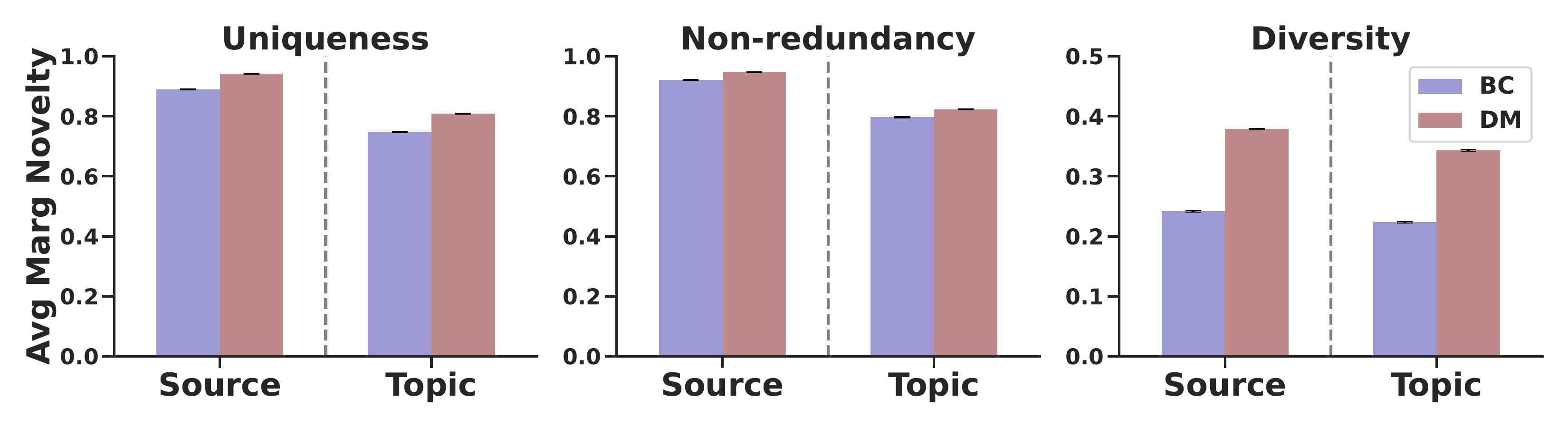}
    \caption{Robustness check for matched users: Comparison of novelty levels between broadcasting (BC) and direct messaging (DM) using the six novelty metrics. Each bar graph represents the average novelty per tie and message for each channel, with error bars delineating the 95\% confidence intervals. Across all metrics, results consistently demonstrate that, on average, the information transmitted through a tie on DM per message is significantly more novel than that transmitted through ties on BC. The results are consistent with the main text.}
    \label{fig:fig1_matching}
\end{figure}

\begin{figure}[H]
    \centering
    \includegraphics[width=0.4\linewidth]{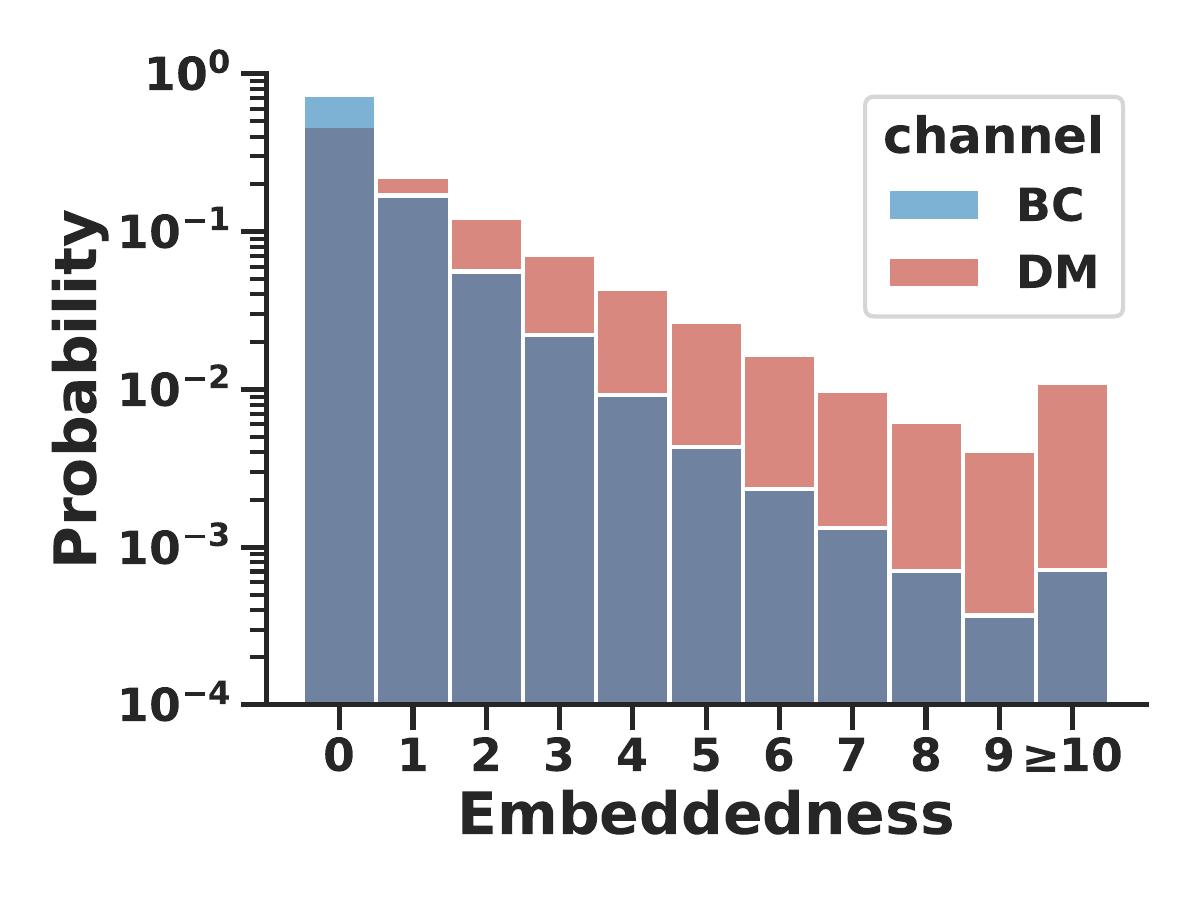}
    \caption{Robustness check for matched users: Distribution of social embeddedness (measured by the number of mutual chat groups for ties in each channel). For both BC and DM, we examine social ties that shared at least one message through that channel. The results are consistent with the main text.}
    \label{fig:hist_matching}
\end{figure}

\begin{figure}[H]
    \centering
    \includegraphics[width=0.7\linewidth]{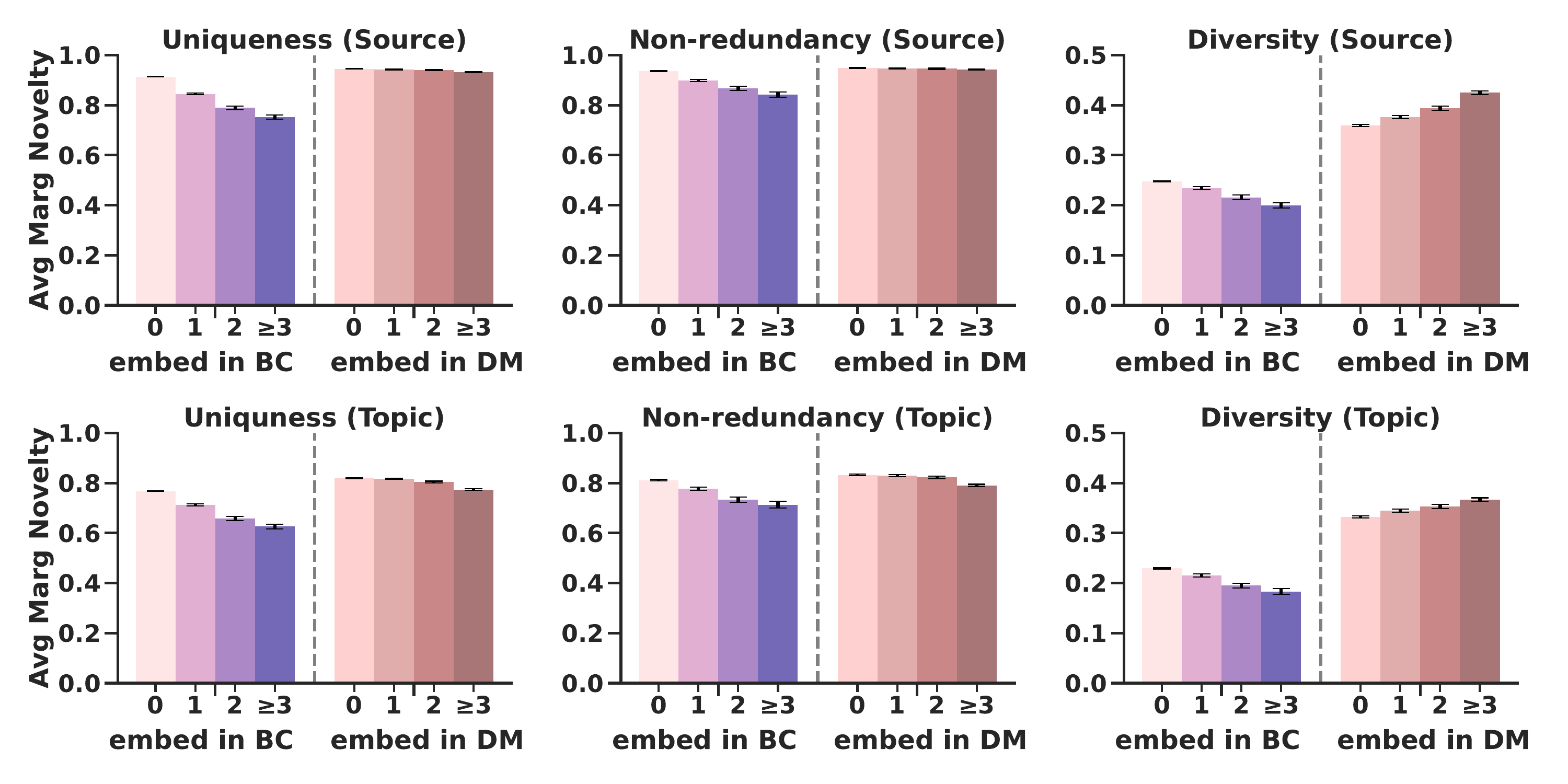}
    \caption{Robustness check for matched users: Comparison of average novelty levels of ties in BC and DM, conditional on the embeddedness of the tie. Error bars denote 95\% confidence intervals. The results are consistent with the main text.}
    \label{fig:fig3_matching}
\end{figure}

To address this concern, we performed user-level matching between DM and BC based on the distribution of bandwidth in DM. For example, for a bandwidth level of $d_{i}$, the DM group has $n^{\text{DM}}_{d_{i}}$ users, and the BC group has $n^{\text{DM}}_{d_{i}}$ users. We then randomly resample users from BC to obtain the same probability (i.e., $\text{prob}_{d_{i}} = \frac{n^{\text{DM}}_{d_{i}}}{\sum_{j}{n^{\text{DM}}_{d_{j}}}}$) in BC as that in DM at the bandwidth level $d_{i}$. This process creates a sample with identical distributions of bandwidth in DM and BC. This matching approach ensures comparability between recipients in DM and BC regarding their bandwidth distributions, making the novelty comparison more valid. The results, depicted in Figure~\ref{fig:fig1_matching} to Figure~\ref{fig:fig3_matching}, are consistent with our main findings and further reinforce our conclusions.

%%%%%%%%%%%%%%%%%%%%%%%%%%%%%%%%%%%%%%%%%%%%%%%%%%%%%%%%%%%%%%%%%%%%%%%%%%%%%%
%%%%%%%%%%%%%%%%%%%%%%%%%%%%%%%%%%%%%%%%%%%%%%%%%%%%%%%%%%%%%%%%%%

%%%%%%%%%%%%%%%%%%%%%%%%%%%%%%%%%%%%%%%%%%%%%%%%%%%%%%%%%%%%%%%%%%
%%%%%%%%%%%%%%%%%%%%%%%%%%%%%%%%%%%%%%%%%%%%%%%%%%%%%%%%%%%%%%%%%%

%%%%%%%%%%%%%%%%%%%%%%%%%%%%%%%%%%%%%%%%%%%%%%%%%%%%%%%%%%%%%%%%%%
%%%%%%%%%%%%%%%%%%%%%%%%%%%%%%%%%%%%%%%%%%%%%%%%%%%%%%%%%%%%%%%%%%

\section{Category Heterogeneity} 
\label{si:category-hetero}
In this analysis, we investigate the distribution of novel information across different topic categories transmitted through strong versus weak ties. Our dataset encompasses 175 topics, which we have organized into 21 categories such as ``society,'' ``education,'' and ``culture'' for more insightful interpretations. These categories are detailed in Table~\ref{si:category-distribution}. To determine the novelty of an article, we compare its category with the recipient's list of favorite categories. This list is derived from the recipient's historical interactions before the observation period and is compiled by the platform's algorithmic engineers and product managers. An article is considered novel if its category does not overlap with any categories in the recipient's list of favorites.

\begin{figure}[H]
    \centering
    \includegraphics[width=0.7\linewidth]{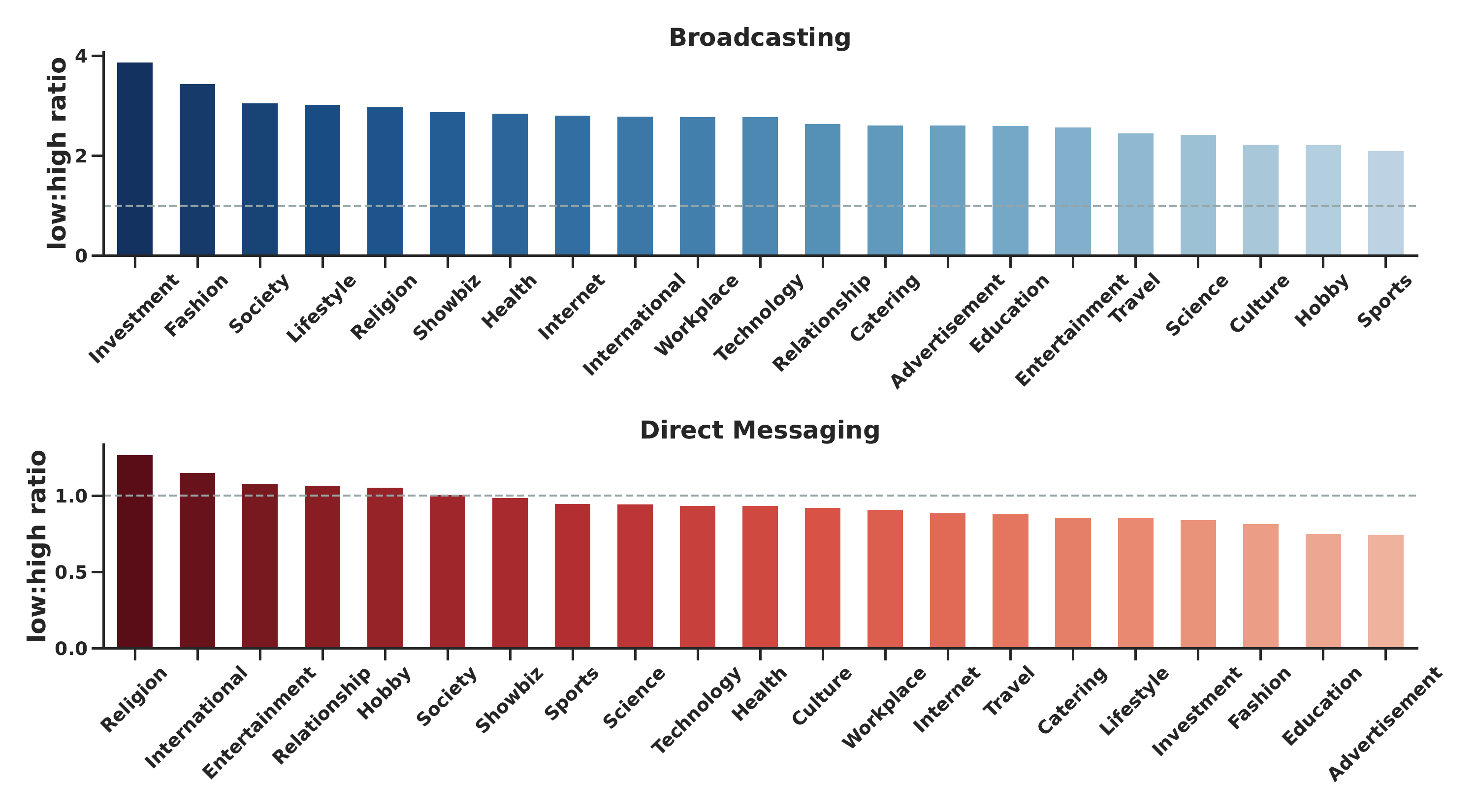}
    \caption{Comparison of strong and weak ties in disseminating novel information across categories}
    \label{fig:category}
\end{figure}

In our analysis, we examine the differences in viral channels across various categories by tracking each instance where a recipient receives an article that is non-redundant to them in terms of category. We measure the ratio of occurrences mediated by low-embedded ties to those mediated by high-embedded ties within the two channels, as illustrated in Figure~\ref{fig:category}. Specifically, we focus on how new categories are disseminated to audiences through high- and low-embedded ties. Low-embedded ties are defined as social ties with no common groups (equal to 0), while high-embedded ties have one or more common groups.
A ratio greater than one for a category in a channel indicates that the category primarily reached its new audience through low-embedded ties. Conversely, a ratio less than one suggests that high-embedded ties were more influential in disseminating the category to new audiences.

%%%%%%%%%%%%%%%%%%%%%%%%%%%%%%%%%%%%%%%%%%%%%%%%%%%%%%%%%%%%%%%%%%
%%%%%%%%%%%%%%%%%%%%%%%%%%%%%%%%%%%%%%%%%%%%%%%%%%%%%%%%%%%%%%%%%%

As illustrated in the upper panel of Figure~\ref{fig:category}, all topics in BC exhibit a ratio significantly greater than one, confirming that the strength of weak (low-embedded) ties manifests for all topics in this channel. This phenomenon is particularly pronounced in the ``investment'' category, which includes interests such as real estate investment, asset management, and stock markets, and in the ``fashion'' category, which covers areas like beauty, jewelry, outfits, and luxury goods.

In DM, the results significantly differ across topics. Most of the ratios fall below one, underscoring the strength of strong (high-embedded) ties in this channel. This observation aligns with our prior findings, suggesting a reduced strength of low-embedded ties in the DM channel. The strength of high-embedded ties is more prominent in the ``advertisement'' category (encompassing articles on coupon information and marketer-generated content), and the ``education'' category (including education philosophy, entrance examination, preschool education, and parenting knowledge).

However, there are a few exceptions in the DM channel where low-embedded ties still play a more important role in diffusing novel information. These exceptions are the ``religion'' category (including Buddhism, Christianity, and Islam), the ``international'' category (covering content on overseas livelihood, immigration, and global military), the ``entertainment'' category (comprising gambling, games, and chess and cards), the ``relationship'' category (including topics on gender and psychological tests), the ``hobby'' category (such as movies, collections, and music), and the ``society'' category (encompassing regional information, people's livelihood, and macroeconomics).

% \bibliographystyle{unsrtnat}
% \bibliography{references}  %%% Uncomment this line and comment out the ``thebibliography'' section below to use the external .bib file (using bibtex) .

%%% Uncomment this section and comment out the \bibliography{references} line above to use inline references.
% \begin{thebibliography}{1}

% 	\bibitem{kour2014real}
% 	George Kour and Raid Saabne.
% 	\newblock Real-time segmentation of on-line handwritten arabic script.
% 	\newblock In {\em Frontiers in Handwriting Recognition (ICFHR), 2014 14th
% 			International Conference on}, pages 417--422. IEEE, 2014.

% 	\bibitem{kour2014fast}
% 	George Kour and Raid Saabne.
% 	\newblock Fast classification of handwritten on-line arabic characters.
% 	\newblock In {\em Soft Computing and Pattern Recognition (SoCPaR), 2014 6th
% 			International Conference of}, pages 312--318. IEEE, 2014.

% 	\bibitem{hadash2018estimate}
% 	Guy Hadash, Einat Kermany, Boaz Carmeli, Ofer Lavi, George Kour, and Alon
% 	Jacovi.
% 	\newblock Estimate and replace: A novel approach to integrating deep neural
% 	networks with existing applications.
% 	\newblock {\em arXiv preprint arXiv:1804.09028}, 2018.

% \end{thebibliography}

\end{document}